# Structural and magnetic phase diagram of CeFeAsO$_{1-x}$F$_x$ and its relationship to high-temperature superconductivity

Jun Zhao**1**, Q. Huang[2], Clarina de la Cruz[1,3], Shiliang Li[1], J. W. Lynn[2], Y. Chen[2,4], M. A. Green[2,4], G. F. Chen[5], G. Li[5], Z. Li[5], J. L. Luo[5], N. L. Wang[5], and Pengcheng Dai[1,3,*]

[1] Department of Physics and Astronomy, The University of Tennessee, Knoxville, Tennessee 37996-1200, USA

[2] NIST Center for Neutron Research, National Institute of Standards and Technology, Gaithersburg, Maryland 20899-6012 USA

[3] Neutron Scattering Science Division, Oak Ridge National Laboratory, Oak Ridge, Tennessee 37831, USA

[4] Department of Materials Science and Engineering, University of Maryland, College Park, Maryland 20742-6393 USA

[5] Beijing National Laboratory for Condensed Matter Physics, Institute of Physics, Chinese Academy of Sciences, Beijing 100080, China

[*] To whom correspondence should be addressed.  E-mail: daip@ornl.gov

**Abstract**

We use neutron scattering to study the structural and magnetic phase transitions in the iron pnictides CeFeAsO$_{1-x}$F$_x$ as the system is tuned from a semimetal to a high-transition-temperature (high-$T_c$) superconductor through Fluorine (F) doping $x$. In the undoped state, CeFeAsO develops a structural lattice distortion followed by a stripe like commensurate antiferromagnetic order with decreasing temperature.  With increasing

Fluorine doping, the structural phase transition decreases gradually while the antiferromagnetic order is suppressed before the appearance of superconductivity, resulting an electronic phase diagram remarkably similar to that of the high-$T_c$ copper oxides. Comparison of the structural evolution of CeFeAsO$_{1-x}$F$_x$ with other Fe-based superconductors reveals that the effective electronic band width decreases systematically for materials with higher $T_c$. The results suggest that electron correlation effects are important for the mechanism of high-$T_c$ superconductivity in these Fe pnictides.

A determination of the structural and magnetic phase transitions in doped transition metal oxides is essential for understanding their electronic properties. For high-transition-temperature (high-$T_c$) copper oxides, the parent compounds are antiferromagnetic (AFM) Mott insulators[1]. When mobile 'electrons' or 'holes' are doped into the parent compounds, the static long-range AFM order is rapidly suppressed and optimal superconductivity emerges after a complete suppression of the static AFM order[2,3]. Much like copper oxide superconductors, high-$T_c$ superconductivity in the recently discovered rare-earth Fe-based oxide systems $R$FeAsO ($R$, rare-earth metal) and (Ba$_{1-x}$K$_x$)Fe$_2$As$_2$ are also derived from either electron[4-8] or hole[9,10] doping of their semimetal parent compounds. Although the parent compound LaFeAsO also exhibits long range static AFM order that is suppressed upon electron doping to induce superconductivity[11-13], there has been no systematic measurement to establish the doping evolution of the AFM order and its relationship to superconductivity. A determination of the structural, magnetic, and superconductivity phase diagram in one of the RFeAsO systems will allow a direct comparison with the phase diagram of high-$T_c$ copper oxides.

Such a comparison is important because it might reveal whether the physics of high-$T_c$ superconductivity in the Fe-based materials is fundamentally related to that of the high-$T_c$ copper oxides[14-16].

In this paper, we report systematic neutron scattering studies of structural and magnetic phase transitions in the Fe pnictides CeFeAsO$_{1-x}$F$_x$ as the system is tuned from a semimetal to a high-$T_c$ superconductor through F doping $x$. We find that CeFeAsO undergoes a structural lattice distortion from tetragonal to orthorhombic structure near 155 K followed by a commensurate AFM ordering on the Fe sublattice below ~140 K as shown in Figs. 1 and 2, similar to that of LaFeAsO (ref. 11). While the structure phase transition temperature decreases gradually with increasing F doping, the AFM ordering temperature and static Fe ordered moment reduce rapidly and vanish before the emergence of superconductivity, resulting an electron phase diagram similar to that of the high-$T_c$ copper oxides (Fig. 1D). Our detailed analysis of the low temperature CeFeAsO$_{1-x}$F$_x$ structures reveal that F doping does not change the Fe-As distance but reduces the Ce-As distance and Fe-As-Fe angles (Fig. 3). The results suggest that the main effect of F doping is to transfer electrons from the Ce-O/F layers to the As-Fe-As block (Fig. 3A). Comparison of the structural evolution of CeFeAsO$_{1-x}$F$_x$ with other rare-earth Fe pnictides[8,11,17,18] and (Ba$_{1-x}$K$_x$)Fe$_2$As$_2$ (refs. 10, 19) suggests that the effective electronic band widths ($W$) in these materials are controlled by the Fe-As-Fe angle, revealing a systematic trend that the parent compounds with smaller band widths tend to have higher $T_c$ upon doping (Fig. 4). The results suggest that although parent compounds of iron pnictides are semimetals, electron correlation effects are very important to their properties[14-16].

We use neutron diffraction to study the structural and magnetic phase transitions in polycrystalline nonsuperconducting CeFeAsO$_{1-x}$F$_x$ with $x$ = 0, 0.02, 0.04, 0.06 (as confirmed by measurements using a commercial SQUID) and superconducting CeFeAsO$_{1-x}$F$_x$ with $x$ = 0.16 (with the onset $T_c$ = 35 K as determined by the susceptibility measurement using a SQUID) using the method described in Ref. 6. Our experiments are carried out on the BT-1 high resolution powder diffractometer and BT-7 thermal triple-axis spectrometer at the NIST Center for Neutron Research, Gaithersburg, Maryland. Some measurements were also performed on the HB-3 thermal triple-axis spectrometer at the High Flux Isotope Reactor, Oak Ridge National Laboratory.

In previous work, it was found that LaFeAsO undergoes a structural distortion below 155 K, changing the symmetry from tetragonal (space group *P4/nmm*) to monoclinic (space group *P112/n*)[11] or orthorhombic (space group *Cmma*)[20], and followed by a long range commensurate AFM order with a stripe like spin structure below ~137 K (ref. 11). For convenience in comparing the low temperature nuclear and magnetic structures, we use orthorhombic *Cmma* space group to describe the low temperature structural data in this paper. Since CeFeAsO$_{1-x}$F$_x$ has rare earth Ce which carries a local magnetic moment[6] and therefore different from the nonmagnetic La in LaFeAsO$_{1-x}$F$_x$ (ref. 11), we first need to determine whether this material has the same lattice distortion and magnetic structure as those of LaFeAsO$_{1-x}$F$_x$. Our high-resolution neutron powder diffraction measurements on BT-1 confirm that the lattice symmetry of CeFeAsO also displays the tetragonal to orthorhombic transition below ~158 K (Figs. 1D and 2A), where the $(2,2,0)_T$ peak in the tetragonal phase is split into $(0,4,0)_O$ and $(4,0,0)_O$ peaks in the orthorhombic phase (inset in Fig. 2A).

To see if the Fe spins in CeFeAsO exhibit the same magnetic order as that of LaFeAsO (ref. 11), we carried out measurements on BT-7. The Ce moments order magnetically below ~4 K (ref. 6 and Fig. 2E), we took data at 40 K to avoid any possible induced-moment influence of Ce on the intensities of the Fe magnetic peaks (Fig. 1C). Comparison of Fig. 1C with the same scan at 170 K (see Supporting Online Material) and with Fig. 3c in ref. 11 for LaFeAsO immediately reveals that the Fe magnetic unit cell in CeFeAsO can be indexed as $\sqrt{2}a_N \times \sqrt{2}b_N \times c_N$, where $a_N$, $b_N$, and $c_N$ are nuclear lattice parameters of the unit cell (see Table 1a). This indicates that CeFeAsO has the same stripe-type in-plane Fe magnetic structure as that of LaFeAsO, but the *c*-axis nearest-neighbor spins are parallel in CeFeAsO rather than anti-parallel as in LaFeAsO. Hence there is no need to double unit cell along the *c*-axis (Fig. 1A), and an excellent fit to the data is achieved using the magnetic and nuclear unit cells in Figs. 1A and 1B as shown by the solid red line of Fig. 1C. The ordered iron moment is 0.8(1) $\mu_B$ at 40 K, where numbers in parentheses indicate uncertainty in the last decimal place and $\mu_B$ denotes Bohr magneton. The magnitude of the Fe moment in CeFeAsO is about twice that of the Fe ordered moment in LaFeAsO (ref. 11). We also determined the Ce magnetic structure using data collected at 1.7 K (see Supporting Online Material) and found a strong coupling between the Fe and Ce moment below 20 K (Figs. 2E-2G). Our determined Ce and Fe magnetic structures are shown in Figs. 1A and 1B. The lack of the *c*-axis unit cell doubling in the Fe magnetic structure is likely due to magnetic interactions of rare earth Ce ions, which doubles the nuclear unit cell along the $c$-axis (Fig. 1A).

Having shown that the lattice distortion and Fe magnetic unit cells are rather similar between CeFeAsO and LaFeAsO, it is important to determine the evolution of the lattice and magnetic structures with increasing F doping as superconductivity is induced. If the stripe-type AFM order in CeFeAsO and LaFeAsO is a spin-density-wave (SDW) instability arising from a nested Fermi surface[21], electron doping will change the electron and hole pocket sizes, but may[23] or may not[24] induce incommensurate SDW order. For pure metallic Cr (ref. 22), where the SDW order has a long wavelength incommensurate magnetic structure, electron/hole doping quickly locks the SDW to commensurate antiferromagnetism with an ordered moment that is doping independent[22]. Figure 2 summarizes the structural and magnetic phase transition temperatures for CeFeAsO$_{1-x}$F$_x$ with $x = 0, 0.02, 0.04, 0.06$. Inspection of Figs. 2A-2D and their insets immediately reveals that the onset lattice distortion temperature (seen as the initial drop in $(2,2,0)_T$ peak intensity) and the magnitude of the lattice distortion (the low temperature splitting of the $(0,4,0)_O$ and $(4,0,0)_O$ peaks) both decrease gradually with increasing $x$ (Fig. 1D). On the other hand, the wavevector positions and coherence-length limits of the $(1,0,2)_M$ magnetic peaks [$Q = 1.838(1), 1.833(1), 1.837(1)$, and $1.831(3)$ Å$^{-1}$; and $\xi = 140(6)$, 137(8), 134(11), and 140(30) Å for $x = 0, 0.02, 0.04, 0.06$, respectively (see inset of Fig. 2G)] are doping independent, and indicate no observable commensurate to incommensurate phase transition. The integrated intensity of the $(1,0,2)_M$ magnetic peak decreases rapidly with increasing $x$ and essentially vanishes near $x = 0.06$ (inset in Fig.

1D). The corresponding Néel temperatures for $T_N$(Fe) and $T_N$(Ce) are determined by measuring the temperature dependence of the $(1,0,2)_M$ magnetic reflection (Figs. 2E-2H). The resulting structural and magnetic phase diagram of $CeFeAsO_{1-x}F_x$ together with superconducting transition temperatures[6] is summarized in Fig. 1D.

Figure 3 summarizes the impact of F-doping on the crystal structure of $CeFeAsO_{1-x}F_x$ obtained from our detailed refinement analysis of the BT-1 data. The undoped CeFeAsO has an orthorhombic low-temperature structure with $c > a > b$ (Fig. 3A). Doping fluorine gradually suppresses both the $a$ (the long Fe-Fe nearest-neighbor distance) and $c$ axes lattice constants while leaving the $b$-axis (the short Fe-Fe nearest-neighbor distance) unchanged (Fig. 3B). The reduction in the $c$-axis lattice constant is achieved via a large reduction of the Ce-As distance, while the Ce-O/F and As-Fe-As block distances actually increase with increasing F-doping (Figs. 3C and 3E). This suggests that the effect of F-doping is to bring the Ce-O/F charge transfer layer closer to the superconducting As-Fe-As block, and thereby facilitating electron charge transfer (Fig. 3A). Since the Fe-As distance (2.405 Å) is essentially doping independent (Fig. 3E), the strong hybridization between the Fe $3d$ and the As $4p$ orbitals[25] is not affected by electron-doping. On the other hand, if we assume that the Fe-Fe nearest-neighbor ($J_1$) and next-nearest-neighbor exchange couplings ($J_2$) are mediated through the electron Fe-As-Fe hopping and controlled by the Fe-As-Fe angles[26], Figure 3D suggests that $J_2$ and one of the nearest-neighbor exchange constants ($J_1$) decrease with increasing F-doping while the other $J_1$ remains unchanged.

In a previous work on the phase diagram of oxygen deficient $R$FeAsO$_{1-\delta}$ (ref. 7), it was found that systematically replacing $R$ from La, to Ce, Pr, Nd, and Sm in $R$FeAsO$_{1-\delta}$ resulted a gradual decrease in the $a$-axis lattice parameters and increase in $T_c$. If $T_c$ for different Fe-based superconductors is indeed correlated to their structural properties, one would expect to find a systematic trend between $T_c$ and the Fe-As-Fe bond angles, since the exchange couplings ($J_1$ and $J_2$) and therefore the electronic band width $W$ are directly related to the Fe-As-Fe bond angles[14,26] (Fig. 4A). Figures 4B and 4C plot the Fe-As(P)-Fe angles and Fe-Fe/Fe-As(P) distances versus maximum $T_c$ for different Fe-based rare-earth oxypnictides[8,10,11,17,18,27,28] and Ba$_{1-x}$K$_x$Fe$_2$As$_2$ (ref. 17) superconductors. While the Fe-Fe/Fe-As(P) distances may not have a clear trend amongst different Fe-based superconductors, it is remarkable that the maximum $T_c$ appears to be directly related to the Fe-As(P)-Fe angles for a variety of materials (Fig. 4C). This suggests that the most effective way to increase $T_c$ in Fe-based superconductors is to decrease the Fe-As(P)-Fe bond angles, and thereby reducing the electronic band width $W$.

In summary, we have mapped out the structural and magnetic phase transitions of CeFeAsO$_{1-x}$F$_x$ and found that the Fe static AFM order vanishes before the appearance of superconductivity[29]. The phase diagram of CeFeAsO$_{1-x}$F$_x$ is therefore remarkably similar to that of the high-$T_c$ copper oxides[1-3]. In addition to suppressing the static antiferromagnetism and inducing superconductivity, F doping also reduces the long-axis of the orthorhombic structure in the undoped CeFeAsO and decreases the Fe-As-Fe bond angles. Comparison of structural parameters of various Fe-based superconductors reveals that the Fe-As(P)-Fe bond angle, and therefore the electronic band width $W$, decreases systematically for superconductors with increasing $T_c$s. If we assume that the effective Coulomb interaction $U$ is comparable amongst different classes of Fe-based superconductors[14], these results suggest that the controlling parameter $U/W$ increases for samples with higher $T_c$s. This means that electron correlations effects, or Mott Physics, becomes increasingly important and should be taken into account as we consider a mechanism for high-$T_c$ superconductivity in these Fe-based materials.

**Acknowledgements** We thank Elbio Dagotto, Adriana Moreo, Randy Fishman, and Thomas Maier for helpful discussions. We also thank J. L. Zarestky for his help on the HB-3 measurements. This work is supported by the US National Science Foundation through DMR-0756568, by the US Department of Energy, Division of Materials Science, Basic Energy Sciences, through DOE DE-FG02-05ER46202. This work is also supported in part by the US Department of Energy, Division of Scientific User Facilities, Basic Energy Sciences. The work at the Institute of Physics, Chinese Academy of Sciences, is supported by the National Science Foundation of China, the Chinese Academy of Sciences and the Ministry of Science and Technology of China.

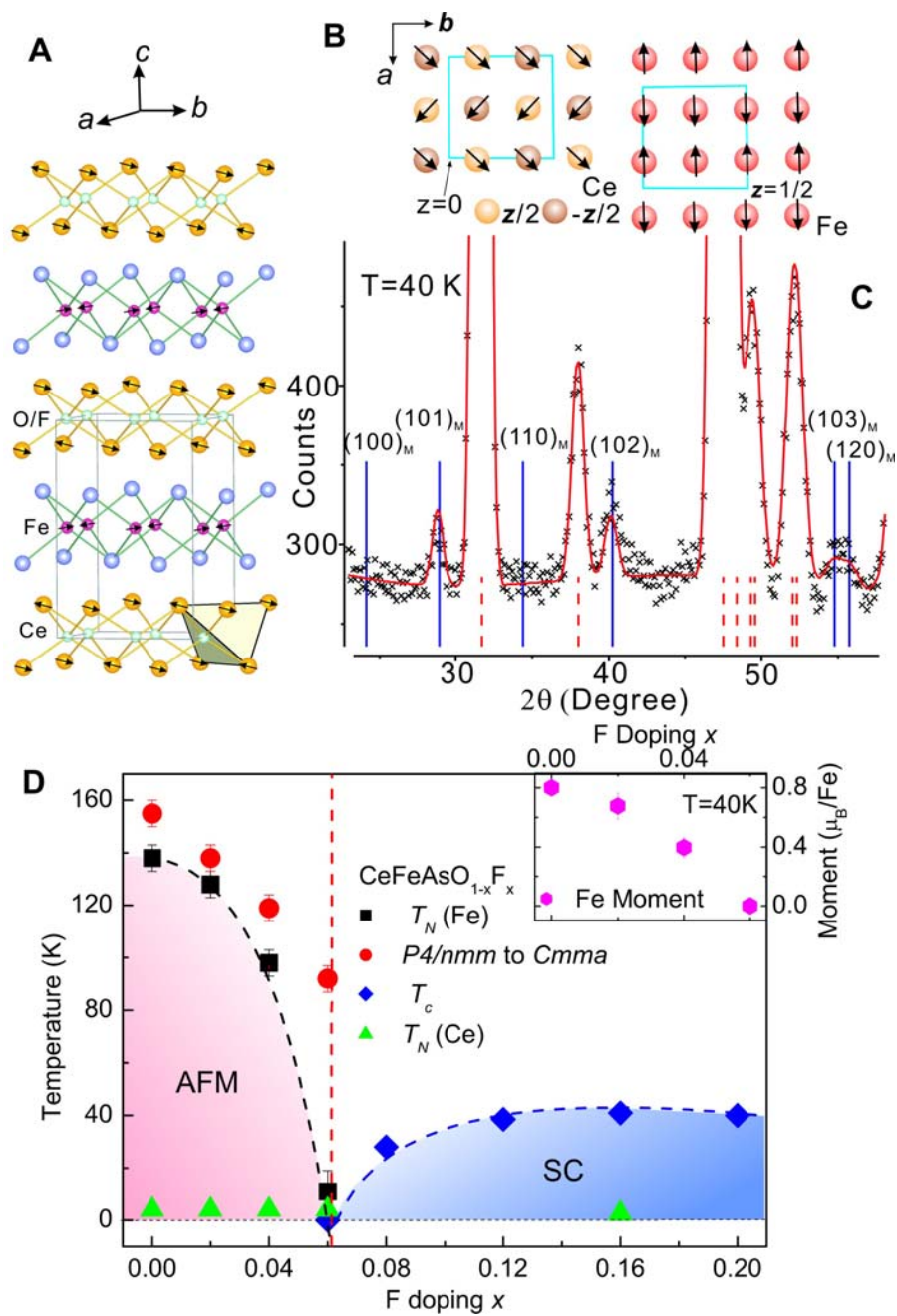

**Figure 1 Low temperature magnetic structures for Ce and Fe in CeFeAsO and the structural and magnetic phase diagram of CeFeAsO$_{1-x}$F$_x$.** The data in panel **C**) were collected using BT-7 with an incident beam wavelength λ = 2.36 Å with pyrolytic graphite (PG) (0,0,2) as monochromator and PG filter. **A**) The three dimensional antiferromagnetic structures of Ce and Fe as determined from our neutron diffraction

data. **B**) The magnetic unit cells of Ce and Fe. The Fe moments lie in the *a-b* plane and form a stripe-type antiferromagnetic structure similar to that of LaFeAsO (ref. 11), while nearest-neighbor spins along the *c*-axis are parallel and so there is no need to double the magnetic cell along the *c*-axis. **C**) Observed (crosses) and calculated (solid line) neutron powder diffraction intensities of CeFeAsO at 40 K using space group *Cmma* for nuclear structure and **A**,**B**) for magnetic structure. The dashed vertical lines indicate the expected nuclear Bragg peak positions while the solid vertical lines represent magnetic Bragg peak positions for the spin structure of the right panel of **B**). **D**) The structural and magnetic phase diagram determined from our neutron measurements on CeFeAsO$_{1-x}$F$_x$ with $x = 0$, 0.02, 0.04, 0.06, 0.16. The red circles indicate the onset temperature of *P4/nmm* to *Cmma* phase transition. The black squares and green triangles designate the Néel temperatures of Fe $T_N$(Fe) and Ce $T_N$(Ce), respectively, as determined from neutron measurements in Figs. 2**E**-2**H**. The superconducting transition temperatures for $x = 0.08$, 0.012, 0.016, 0.20 are from the onset $T_c$ of the resistivity measurements adapted from ref. 6. We note that $T_c$ determined from susceptibility measurements in Fig. 4 has a lower value. The inset in **D**) shows the F doping dependence of the Fe moment as determined from the intensity of the $(1,0,2)_M$ magnetic peak at 40 K, where the influence of the Ce moment on the Fe magnetic Bragg peak intensity can be safely ignored.

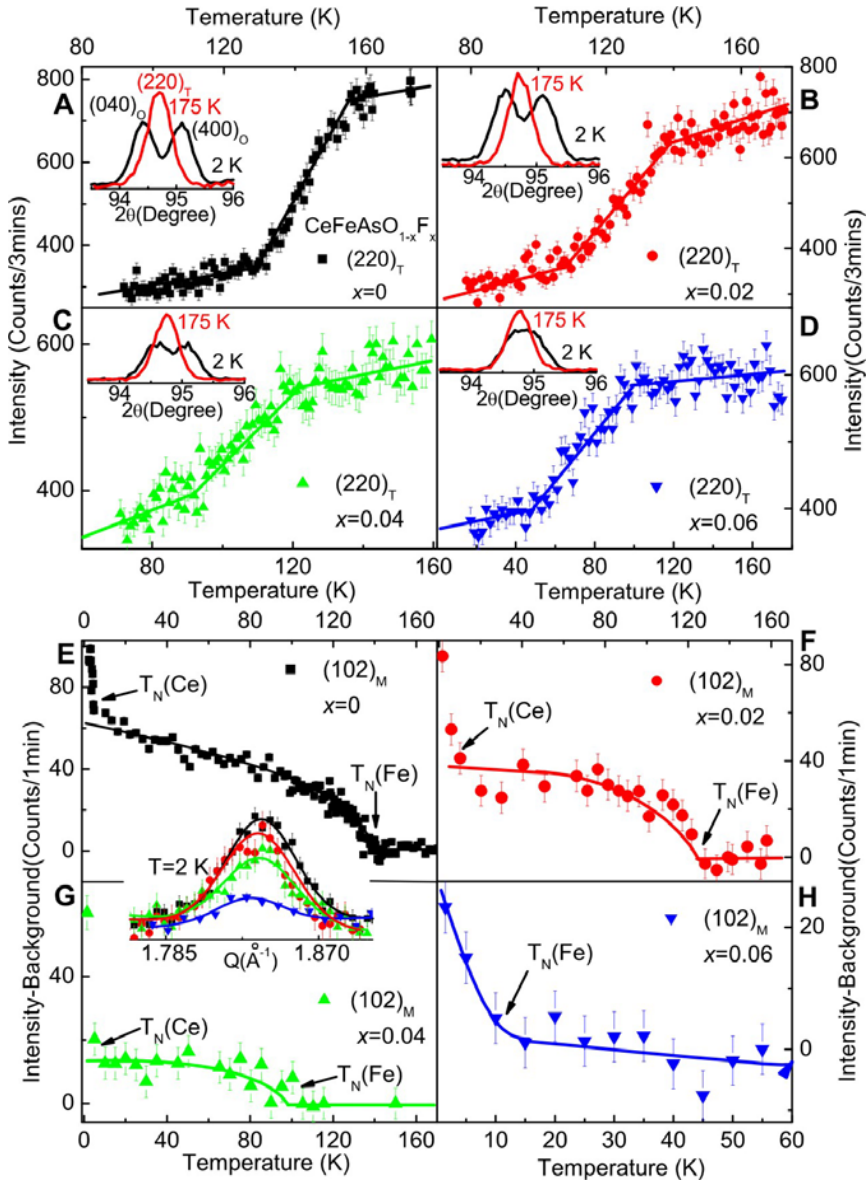

**Figure 2 Structural and magnetic phase transition temperatures as a function of increasing F doping in CeFeAsO$_{1-x}$F$_x$.** The data in **A-D**) and **E-G**) were collected on BT-1 and BT-7, respectively. The *Q*-scan for $x = 0.06$ (inset in **E**) and **H**) were carried out on HB-3 using similar setup as BT-7. The BT-1 diffractometer has a Ge(3,1,1) monochromator and an incident beam wavelength of $\lambda = 2.0785$ Å. **A-D**) Temperature dependence of the $(2,2,0)_T$ (T denotes tetragonal) nuclear reflection indicative of a structural phase transition[11] for various *x*. The insets show the $(2,2,0)_T$ reflection above

and below the transition temperatures[11]. **E-H**) Temperature dependence of the order parameter at the magnetic Bragg peak position $(1, 0, 2)_M$ as a function of F doping. The large increase in intensity below 4 K is due to Ce ordering, as confirmed by temperature dependence of the Ce-only magnetic Bragg peak $(0,0,1)_M$ (see Supporting Online Material). The inset shows the doping dependence of the $(1,0,2)_M$ Bragg peak normalized to the nuclear Bragg peak intensity. The peak positions and widths are essentially doping independent, suggesting that the AFM order is commensurate at all doping levels.

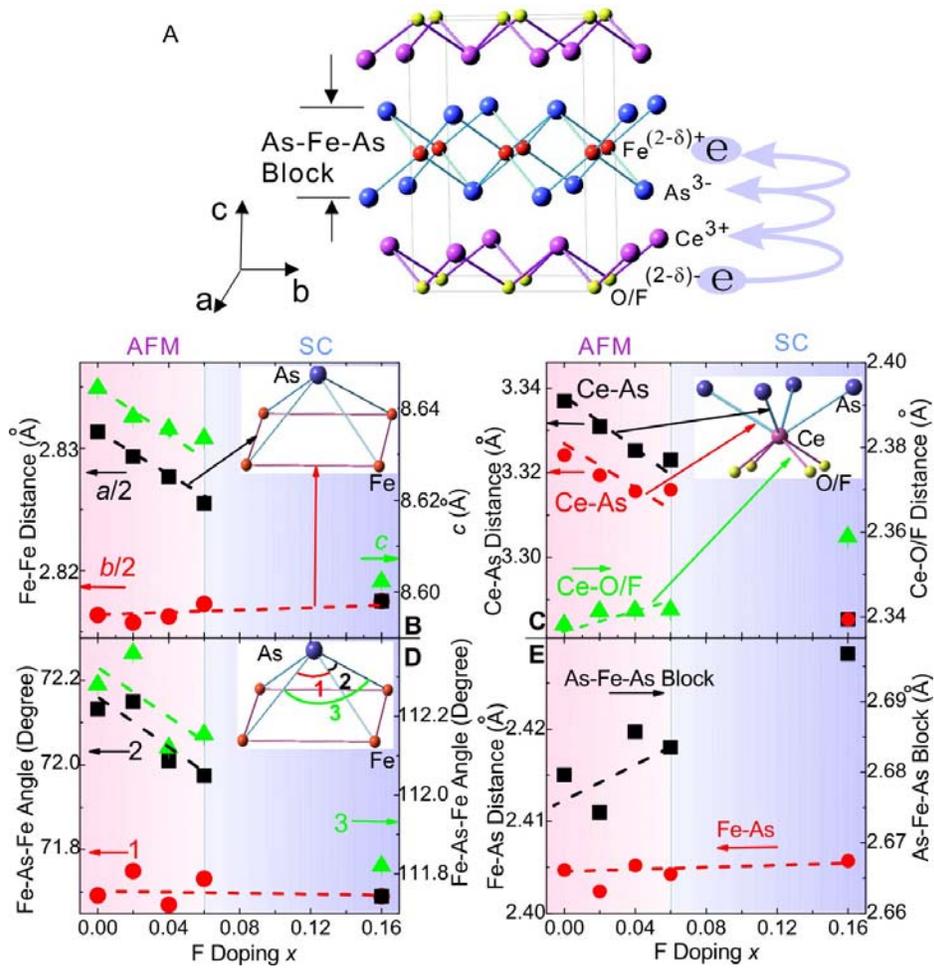

**Figure 3 Low temperature structural evolution of $CeFeAsO_{1-x}F_x$ as a function of F doping obtained from analysis of the BT-1 data.** The atomic positions of $CeFeAsO_{1-x}F_x$ are shown in Table 1b and the effect of F doping is to expand the Fe-As-Fe block and to move the Ce-O/F block closer to Fe-As-Fe block, thereby facilitating electron doping to the superconducting Fe-As-Fe layer. **A)** schematic diagram defining the Fe-As-Fe block and illustrating the process of electron doping. **B)** $a$, $b$, $c$ lattice constants of the orthorhombic unit cell and the two Fe-Fe nearest-neighbor distances as a function of F doping. **C)** Ce-O/F and Ce-As distances as a function of F doping. The slight increase in the Ce-O/F block size is compensated by much larger reduction in the Ce-As distance, resulting an overall $c$-axis lattice contraction as shown in **B**). **D)** Fe-As-Fe bond angles as defined in the inset versus F doping. While angle 1 hardly changes with doping, angles 2 and 3 decrease substantially with increasing F doping. **E**) The Fe-As bond distance and As-Fe-As block size versus F doping. The Fe-As distance is independent of F doping.

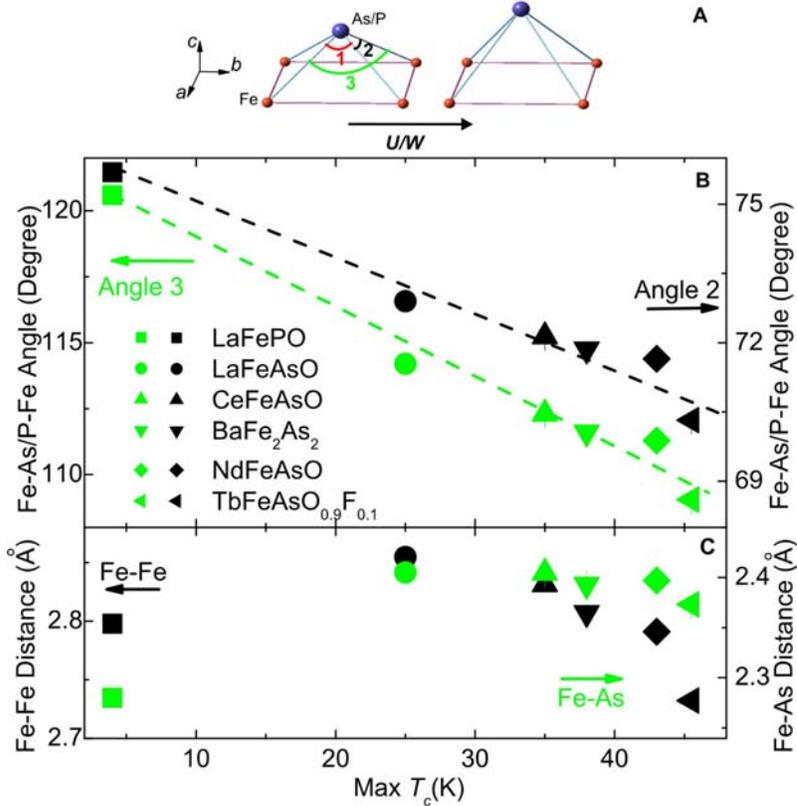

**Figure 4. Fe-As(P)-Fe bond angles, Fe-Fe, and Fe-As(P) distances for different Fe-based superconductors.** There is a systematic decrease in the Fe-As(P)-Fe bond angle for Fe-based superconductors with higher $T_c$, suggesting that electron correlation effects are important. **A**) Schematic illustration of what happens to the Fe-As-Fe tetrahedron for Fe-based superconductors. The smaller Fe-As(P)-Fe angle on the right results in a narrower electronic band width $W$, increasing $U/W$, and thereby raising $T_c$. **B,C**) Dependence of the maximum-$T_c$ on Fe-As(P)-Fe angle and Fe-Fe/Fe-As(P) distance. The Fe-As(P)-Fe angles and Fe-Fe/Fe-As(P) distances are computed using atomic positions

given in Refs. 18, 27 for LaFePO; Ref. 11 for LaFeAsO; present paper for CeFeAsO; Ref. 10 for $BaFe_2As_2$; Ref. 28 for NdFeAsO; and Ref. 8 for $TbFeAsO_{0.9}F_{0.1}$. Note here we used the maximum-$T_c$ obtained from susceptibility measurement, which is lower than that of the resistivity measurement on the same system.

**Table 1a**. Refined structure parameters of $CeFeAsO_{1-x}F_x$ with $x = 0$ at 175 K and $x = 0.16$ at 60 K. Space group: *P4/nmm*.
CeFeAsO, $a = 3.99591(5)$, $c=8.6522(1)$ Å; $CeFeAsO_{0.84}F_{0.16}$, $a = 3.98470(3)$, $c = 8.6032(1)$ Å.

| Atom | site | x | y | z (x = 0) | B (Å²)(x = 0) | z (x = 0.16) | B (Å²)(x = 0.16) |
|---|---|---|---|---|---|---|---|
| Ce | 2c | ¼ | ¼ | 0.1413(3) | 0.34(4) | 0.1480(4) | 0.58(5) |
| Fe | 2b | ¾ | ¼ | ½ | 0.25(4) | ½ | 0.09(3) |
| As | 2c | ¼ | ¼ | 0.6546(2) | 0.28(3) | 0.6565(3) | 0.27(4) |
| O | 2a | ¾ | ¼ | 0 | 0.30(5) | 0 | 0.50(4) |

$x = 0$, $Rp = 5.02\%$, $wRp = 6.43\%$, $\chi^2 = 1.336$;
$x = 0.16$, $Rp = 5.94\%$, $wRp = 8.24\%$, $\chi^2 = 2.525$.

**Table 1b**. Refined structure parameters of $CeFeAsO_{1-x}F_x$ with x = 0, 0.02, 0.04, 0.06 at 1.4 K. Space group: *Cmma*. Atomic positions: Ce: 4g (0, ¼, z); Fe: 4b (¼, 0, ½), As: 4g (0, ¼, z), and O/F: 4a (¼, 0, 0).

| Atom | | x = 0 | x = 0.02 | x = 0.04 | x = 0.06 |
|---|---|---|---|---|---|
| | a (Å) | 5.66263(4) | 5.65865(9) | 5.6553(1) | 5.6511(1) |
| | b (Å) | 5.63273(4) | 5.63155(9) | 5.6325(1) | 5.6346(1) |
| | c (Å) | 8.64446(7) | 8.6382(1) | 8.6355(2) | 8.6335(1) |
| Ce | z | 0.1402(2) | 0.1417(4) | 0.1419(4) | 0.1420(3) |
| | B (Å²) | 0.36(2) | 0.37(6) | 0.31(6) | 0.46(5) |
| Fe | B (Å²) | 0.34(2) | 0.38(4) | 0.30(3) | 0.34(3) |
| As | z | 0.6553(1) | 0.6548(3) | 0.6555(3) | 0.6554(2) |
| | B (Å²) | 0.45(2) | 0.50(6) | 0.36(5) | 0.24(4) |

| O/F | $B$ (Å$^2$) | 0.54(2) | 0.53(6) | 0.64(6) | 0.63(5) |
|---|---|---|---|---|---|
| | $Rp$ (%) | 4.31 | 5.44 | 4.90 | 4.71 |
| | $wRp$ (%) | 5.60 | 6.72 | 6.31 | 6.16 |
| | $\chi^2$ | 2.192 | 1.258 | 0.966 | 0.9622 |

**Supporting Online Material**

# Structural and magnetic phase diagram of CeFeAsO$_{1-x}$F$_x$ and its relationship to high-temperature superconductivity

Jun Zhao, Q. Huang, Clarina de la Cruz, Shiliang Li, J. W. Lynn, Y. Chen, M. A. Green, G. F. Chen, G. Li, Z. Li, J. L. Luo, N. L. Wang, and Pengcheng Dai

Figure S1 below shows the BT-7 measurements on CeFeAsO at 1.5 K, 40 K, and 160 K. The $(0,0,1)_M$ peak corresponds to a doubling of the $c$-axis unit cell for the Ce magnetic order, which disappears above the Ce Néel temperature of ~4 K. Comparison of the 40 K and the 160 K scans reveals the positions of the Fe magnetic peaks as shown in Fig. 1C. Our detailed refinements for the determination of the Ce magnetic structure and its interaction with the iron spins will be published separately.

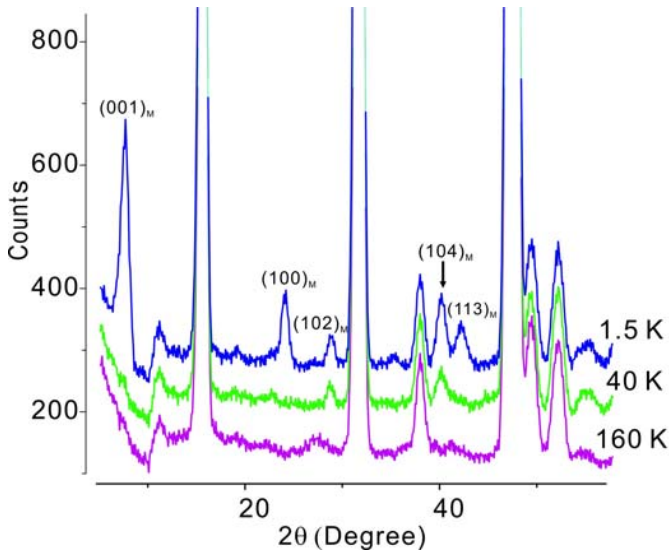